\documentclass{mem}
\usepackage{natbib}
\usepackage{balance}
\usepackage{graphicx}
\usepackage{flushend}
\usepackage[breaklinks]{hyperref}
\usepackage{xcolor}
\usepackage{chemformula}
\usepackage{newtxmath}
\usepackage{amsmath}
\usepackage{txfonts}
\idline{75}{282}

\begin{document}
\def\teff{$T\rm_{eff }$}
\def\kms{km~s$^{-1}$}

\newcommand{\asec}{$^{\prime\prime}$}

\title{
Astrochemistry on Galactic scales
}

\author{
L. \,Colzi\inst{1} 
\and V. M. \,Rivilla\inst{1}
\and M. T. \,Beltr\'an\inst{2}
\and C. Y. \,Law\inst{3,4}
\and E. \,Redaelli\inst{5}
\and \\ M. \,Padovani\inst{2}
          }

\institute{
Centro de Astrobiología (CSIC-INTA), Ctra Ajalvir km 4, 28850 Torrejón de Ardoz, Madrid, Spain
\email{lcolzi@cab.inta-csic.es}
\and
INAF-Osservatorio Astrofisico di Arcetri, Largo E. Fermi 5, 50125
Firenze, Italy
\and 
European Southern Observatory, Karl-Schwarzschild-Strasse 2, 85748 Garching, Germany
\and
Department of Space, Earth \& Environment, Chalmers University of Technology, 412 96 Gothenburg, Sweden
\and
Max-Planck-Institut für extraterrestrische Physik, Gießenbachstraße 1, 85748 Garching bei München, Germany
}

\authorrunning{Colzi et al.}

\titlerunning{Galactic astrochemistry}

\date{Received: 29-02-2024; Accepted: 07-07-2024}

\abstract{The increasing number of observations towards different environments in the Milky Way, as well as theoretical and experimental works, are improving our knowledge of the astrochemical processes in the interstellar medium (ISM). In this chapter we report some of the main projects to study the chemical complexity and isotopic ratios across the Galaxy. High-sensitivity spectral surveys covering broad bandwidths towards Galactic Center molecular clouds (e.g. G+0.693-0.027) and star-forming regions (e.g. the hot core G31.41+0.31) are revealing very rich astrochemical reservoirs, which include molecules of prebiotic interest. At the same time, isotopic ratios (e.g. $^{12}$C/$^{13}$C and $^{14}$N/$^{15}$N) can give important information on the Galactic chemical evolution, as well as on chemical local processes due to the physical conditions of the molecular clouds. We also highlight the role of cosmic rays as a key agent affecting the interstellar chemistry described above. 

}

\maketitle{}

\section{Introduction}

Astrochemistry is now living in a golden age. As of February 2024, more than 305 molecules have been detected in the interstellar medium (ISM) or circumstellar shells, of which $\sim$90 from 2020, without including isotopologues.
In recent years, the improved sensitivity and broadband capabilities of current telescopes (e.g. GBT 100m, Yebes 40m, IRAM 30m, and ALMA) have allowed the discovery of molecular species with increasing complexity. 
Many observational campaigns have been carried out to study the molecular complexity within the whole Milky Way, from its center  to the outer parts. 
%
%
In the Galactic Center, the Sgr B2(N) hot molecular core and surrounding envelope have provided the first detections in space of many species, including proposed precursors of amino acids (amino acetonitrile, \ch{NH2CH2CN}; \citealt{belloche2008}); branched species (iso-propyl cyanide, \ch{i-C3H7CN}, \citealt{belloche2014}); and chiral molecules (propylene oxide, \ch{CH3CHCH2O}, \citealt{mcguire2016}). Towards the neighbouring molecular cloud G+0.693-0.027 many molecular RNA precursors have first been detected, such as hydroxylamine (\ch{NH2OH}; \citealt{rivilla2020b}), ethanolamine (\ch{NH2CH2CH2OH}, \citealt{rivilla2021a}), 1-2-ethendiol (\ch{(CHOH)2}, \citealt{rivilla2022a}). Many more chemical studies will be done in the Galactic Center in the context of the ACES (The ALMA CMZ Exploration Survey) Large Program (ID: 2021.1.00172.L; PI: Steven Longmore).
%
%

In the Galactic Disk, the GOTHAM and QUIJOTE ultra-deep spectral surveys 
have demonstrated that the dark cloud TMC-1 exhibits an extremely rich chemistry, including aromatic species, such as benzonitrile (\citealt{mcguire2018}), indene (\citealt{cernicharo2021}; \citealt{burkhardt2021}), or cyanonaphtalene (\citealt{mcguire2021}), among many others.
Towards star-forming regions, detailed studies of archetypical low-mass and high-mass protostellar environments, such as IRAS 16293-2422 B (\citealt{jorgensen2018}) and the G31.41+0.31 hot core (\citealt{mininni2020}), respectively, have provided a complete view of rich chemical reservoir.
%
%
And recently, complex chemistry has been revealed in sources located in the outer edge of the Galaxy, at galactocentric distances $>$12 kpc (\citealt{shimonishi2021,fontani2022a,fontani2022b}).

Within the molecules detected in the ISM, there are also many isotopologues, i.e. molecules in which one of the atoms is substituted by a less abundant isotope. Isotopic ratios of elements, such as $^{14}$N/$^{15}$N, can probe the Milky Way chemical evolution, and can be used as indicators of stellar nucleosynthesis (e.g. \citealt{romano2017,romano2019}). Molecular isotopic ratios are also affected by fractionation processes, which favour the rarer isotopic substitution over the most abundant isotopologue, depending on the local physical conditions of the cloud (e.g. density, temperature, UV field), shaping the measured ratios from the initial one produced by nucleosynthesis. 

The chemical processes that synthesize the molecules and their isotopologues in the ISM are affected by the presence of cosmic rays. 
In particular, cosmic rays with energy below 1~GeV influence the thermochemistry of the shielded molecular gas, which results from the ionisation of both atomic and molecular hydrogen. After the ionisation of H$_{2}$, H$_{2}^{+}$ readily reacts with nascent H$_{2}$ to form the trihydrogen cation H$_{3}^{+}$ . This initiates a series of reactions that lead to the formation of more and more complex species up to prebiotic molecules \citep[see e.g.][for a review]{PadovaniGaches2024}.


In this chapter, we will focus on the most recent discoveries regarding chemical complexity in the Galactic Center molecular cloud G+0.693-0.027 and the hot core G31.41+0.31 (Sect.~\ref{sec-galaxy}), isotopic ratios measured across the Galaxy (Sect.~\ref{sec-ratios}), and we will highlight the importance of cosmic rays for astrochemical processes (Sect.~\ref{sec-cr}).


\vspace{-0.3cm}

\section{Chemical complexity in the Galaxy}
\label{sec-galaxy}

\subsection{A Galactic Center molecular cloud: G+0.693-0.027}

\label{sec-g0693}

The central molecular zone (CMZ, inner 300 pc of the Galaxy) contains 80\% of the dense molecular gas in the Galaxy, but the star formation rate is one order of magnitude lower than in the disk ($\sim$0.1 M$_{\odot}$/yr vs. 1.5--2 M$_{\odot}$/yr, \citealt{longmore2013,barnes2017}). 
The CMZ presents extreme conditions, like a high level of turbulence due to the large internal cloud velocity dispersion ($\sim$15-50~\kms), and widespread high kinetic temperatures (from 50 up to $>$100 K; e.g. \citealt{krieger2017}), which could prevent star formation. 

The G+0.693-0.027 molecular cloud  (hereafter G+0.693) is located $\sim$55\asec\ towards the NE of the Sgr B2(N) hot core. 
The source presents typical linewidths of 15--25 \kms, a density of $\sim$10$^{4}$ cm$^{-3}$ and $T_{\rm kin}$$>$100 K (e.g. \citealt{zeng2018}).
G+0.693 does not show any signposts of ongoing star formation, such as ultracompact HII regions, H$_{2}$O masers, or dust continuum point sources (\citealt{ginsburg2018}). G+0.693 is probably affected by low-velocity shocks as a consequence of a large-scale cloud–cloud collision (\citealt{zeng2020}), likely responsible for its rich chemistry due to the sputtering of molecules from dust grains (see below). The molecular emission from this region is sub-thermally excited because of its low density, and presents excitation temperatures of 5--20 K (much lower than the $T_{\rm kin}$ of 50--150 K). This makes the spectrum much less crowded of lines since only those with low energies are excited, and allow for the identification of many molecular species without contamination as in hot cores (e.g. \citealt{martin2008,zeng2018,rivilla2019,rivilla2020b, jimenez-serra2020, rivilla2021a, jimenez-serra2022, sanandres2023}).

\citet{colzi2022a} studied the D/H ratios of HCN, HNC, HCO$^{+}$, and N$_{2}$H$^{+}$ towards G+0.693 using multiple rotational transitions. The authors discovered the presence of two line components: (i) a turbulent one with a linewidth of 20 \kms\;typical of the GC and already known from other studies for this source (broad component), (ii) a new less turbulent component with a linewidth of 9 \kms, which is very clear from the high-$J$ transitions of the molecules studied (narrow component). \citet{colzi2022a} estimated for the broad component a temperature, $T_{\rm kin}$, of 100 K and H$_{2}$ densities of 0.3--3$\times$10$^{4}$ cm$^{-3}$, and for the narrow component $T_{\rm kin}\leq$30 K and  H$_{2}$ densities increased by at least one order of magnitude to 0.05--1$\times$10$^{6}$ cm$^{-3}$. This is the first indication that a substantial fraction of gas in the GC (10\% in column density) has the density and temperature ideal to form the new generation of stars.

Regarding the chemical complexity in G+0.693, \citet{requena-torres2006, requena-torres2008} and \citet{zeng2018} already showed that it is rich in complex O- and N-bearing species, respectively. 
In recent years, to fully understand the chemical complexity that can be reached in G+0.693, a broadband ultra-high sensitivity spectral survey has been carried out, using the GBT, Yebes 40m, IRAM 30m and APEX telescopes. These observations cover frequencies from $\sim$12.7 up to $\sim$276 GHz, reaching sensitivities down to sub-mK levels (e.g. \citealt{rivilla2023}). Remarkably, this survey has provided the discovery of 16 new interstellar species since 2019. These molecules are members of increasing complexity of different chemical families that contain the five chemical elements essential for life (CHONPS):
CH-bearing species, such as \ch{(CH3)2CCH2} (\citealt{fatima2023});
O-bearing species (HCOCOOH, \citealt{sanz-novo2023}; \ch{(CHOH)2}, \citealt{rivilla2022a}; or \ch{C3H7OH}, \citealt{jimenez-serra2022}); 
N-bearing species (HNCN, \citealt{rivilla2021b}; \ch{C2H3NH2}, \citealt{zeng2021}; or HCCCHNH, \citealt{bizzocchi2020});
NO-bearing species (\ch{NH2OH}, \citealt{rivilla2020b};  \ch{C2H5NCO}, \citealt{rodriguez-almeida2021b}; or \ch{NH2CH2CH2OH}, \citealt{rivilla2021a}); 
S-bearing species (HCOSH, \citealt{rodriguez-almeida2021a}; or HOCS$^+$, \citealt{sanz-novo2024});
and P-bearing species (PO$^{+}$, \citealt{rivilla2022b}). 
Many of these new molecules have been proposed as key molecular precursors of RNA nucleotides in prebiotic experiments that mimic the conditions of early Earth (\citealt{powner2009,patel2015,becker2019}), which suggests that interstellar chemistry could have played a relevant role in the origin of Life in our planet.

\vspace{-0.3cm}

\subsection{A Galactic Disk star-forming region: the G31.41+0.31 hot molecular core}
\label{sec-G31}

\begin{figure*}[h!]
\centering
\resizebox{11cm}{!}{\includegraphics[clip=true]{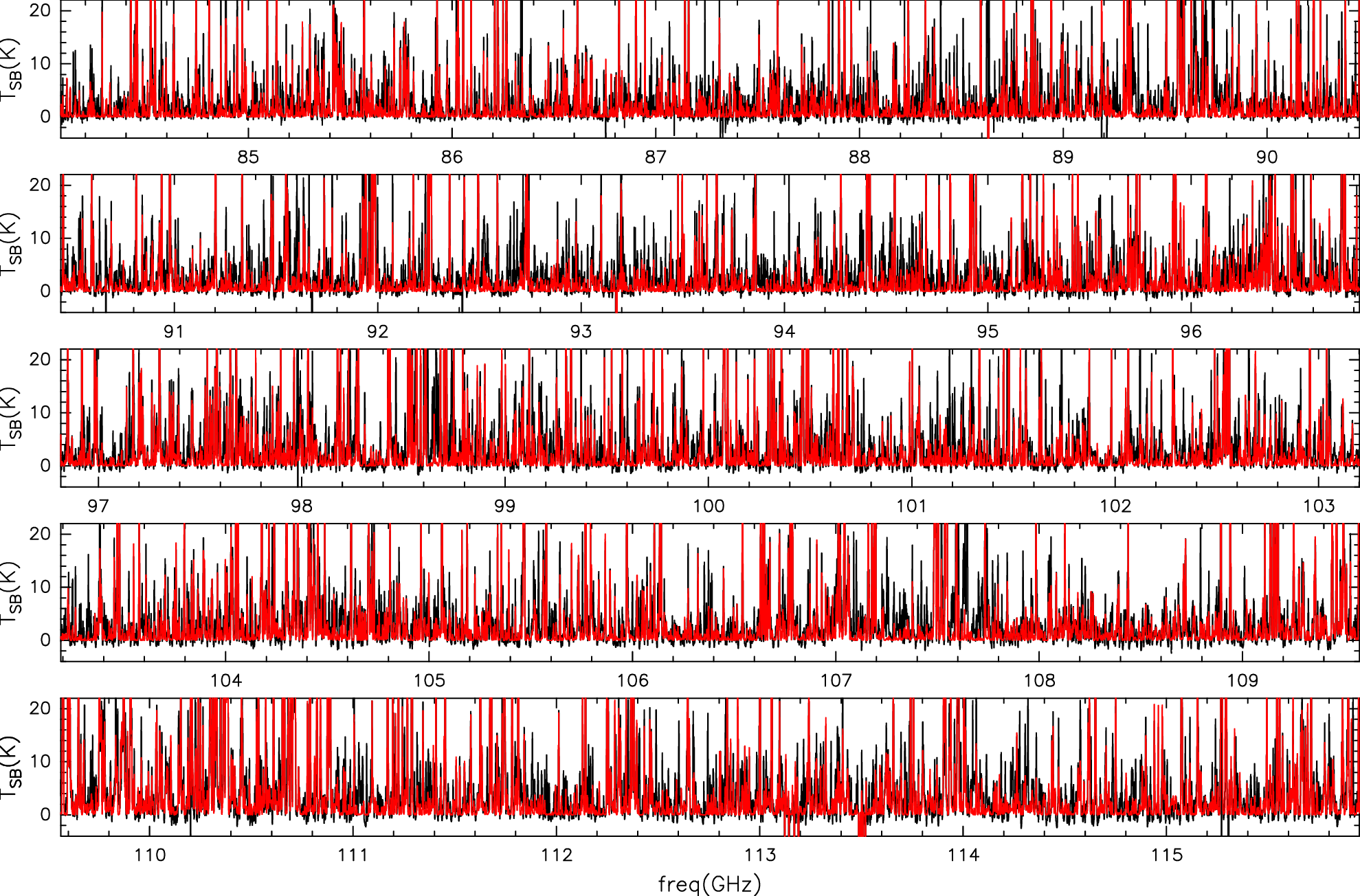}}
\caption{\footnotesize Full ALMA spectrum (black histogram) of the G31.41+0.31 hot core from 84 to 116,  GHz, adapted from \citet{mininni2020} and \citet{colzi2021}. The red solid line is the best fit for all the species detected.}
\label{fig-g31-lines}
\end{figure*}


G31.41+0.31 (hereafter G31) is a well-known hot molecular core located at a distance of 3.75 kpc, with a luminosity $\sim$5$\times$10$^{4}\,L_\odot$ (\citealt{beltran2005}). 
Interferometric observations of CH$_3$CN and other complex organic molecules (COMs) have revealed a velocity gradient in the core, interpreted as the rotation of a toroid (\citealt{beltran2005, girart2009, beltran2018}), as well as evidence of infall (e.g., \citealt{beltran2018, beltran2021}), while SiO observations have detected several outflows associated with the core (\citealt{beltran2018, beltran2021}).
ALMA and VLA high-resolution observations have resolved the dust continuum and free-free emission into at least four sources, indicating that G31 is hosting a massive protocluster (\citealt{cesaroni2010, beltran2021}).

G31 is very chemically rich core (e.g., \citealt{beltran2005, beltran2009, rivilla2017}), presenting prominent emission in a large number of COMs.
Indeed, the first detection of glycolaldehyde (CH$_{2}$(OH)CHO) outside the GC was made towards G31 (\citealt{beltran2009}).
Recently, the ``G31 Unbiased ALMA sPectral Observational Survey'' (GUAPOS, \citealt{mininni2020}; ID: 2017.1.00501.S; PI: Maite Beltrán) has studied the molecular emission across the whole ALMA Band 3,  from $\sim$84\,GHz up to $\sim$116\,GHz, with an angular resolution of 1\asec.2 (4500 au). 
Figure~\ref{fig-g31-lines} shows the  spectrum towards the center of G31, adapted from \citet{mininni2020} and \citet{colzi2021}.
The GUAPOS project has allowed to identify so far more than 40 molecules in the region (and multiple isotopologues), including 23 COMs (\citealt{mininni2020,mininni2023,colzi2021,fontani2023,lopez-gallifa2024}).
%
%
%
A summary of the main results is presented below:

$\bullet$  
\citet{mininni2020} analyzed the emission of three C$_2$H$_4$O$_2$ isomers, that is, methyl formate (CH$_3$OCHO), acetic acid (CH$_3$COOH), and glycolaldehyde (CH$_2$OHCHO), and observed that it is very compact and associated with the inner part of the core.  The abundances of methyl formate and acetic acid toward G31 are higher than those detected toward the well-studied prototype of hot cores, such as Sgr B2(N), and of hot corinos, such as IRAS IRAS 16293-2422B. The comparison of the abundances of the three isomers with those predicted by chemical models suggests the necessity of grain-surface routes, at least for the formation of CH$_3$OCHO see e.g. \citealt{garrod2013,balucani2015,vasyunin2017,coutens2018,skouteris2018}).
%

$\bullet$ Among COMs, those containing the NCO backbone (peptide-bond) are of great interest since peptide-bonds can link amino acids to form proteins (e.g. \citealt{pascal2005}). \citet{colzi2021} studied peptide-like bond molecules, such as isocyanic acid (HNCO), formamide (HC(O)NH$_{2}$), methyl isocyanate (CH$_{3}$NCO), and also more complex species such as acetamide (CH$_{3}$C(O)NH$_{2}$) and N-methylformamide (CH$_{3}$NHCHO). These molecules have been observed together for the first time in the disk of our Galaxy, outside the Galactic Center. The results from this work suggest that these molecules have likely been formed on the surface of interstellar grains during the earliest (and cold) stages of star formation. Once the protostars are born, they start to heat their surroundings and drive molecular outflows. These physical processes are responsible for the evaporation and/or the sputtering of the molecules from the surface of the dust grains.

$\bullet$ The analysis of nine O-bearing and six N-bearing COMs by \citet{mininni2023} indicated that their abundances range from 10$^{-10}$ to 10$^{-6}$. From the comparison of the abundances of these species with those estimated for other twenty-seven sources, it is evident that there is not a unique template for the abundances of COMs in hot molecular cores.  This could be the result of the different physical properties of the sources and their different evolution with time, which suggests the importance of the thermal history for their chemistry.

$\bullet$ \citet{fontani2023} have detected P-bearing molecules in positions separated from the hot core and located along the outflows identified in G31 in previous studies. PN is clearly detected, while PO is tentatively detected. The fact that PN is tightly associated with SiO and other typical shock tracers, such SO, and the lack of a clear detection of PN towards the hot core provides a robust confirmation that PN is likely a product of shock chemistry, as previously shown by \citet{lefloch2016} and \citet{rivilla2020a}, and allows us to rule out formation pathways in hot gas. 

$\bullet$ \citet{lopez-gallifa2024} have analyzed more than 50 species, including oxygen, nitrogen, sulfur, phosphorus, and chlorine species and have compared their abundances with those observed in other chemically-rich sources that represent the initial and last stages of the formation of stars and planets: the hot corino in the Solar-like protostar IRAS 16293$–$2422 B, and the comets 67P/Churyumov- Gerasimenko and 46P/Wirtanen. The comparative analysis reveals that O- and N-bearing molecules exhibit a good correlation for all sources, suggesting a chemical heritage of these species during the process of star formation, and hence an early phase formation of the molecules. On the other hand, S- and P-bearing species do not follow the same correlation. While they are less abundant in the gas phase of star-forming regions, likely because they are predominantly trapped on the surface of icy grains, the cosmic abundances are recovered in the coma of the comet.

\vspace{-0.3cm}
\section{Isotopic ratios in the Galaxy}
\label{sec-ratios}

Isotopic ratios measured using molecules and their isotopologues in molecular clouds
depend on the chemical evolution of the Galaxy due to stellar nucleosynthesis, and also on local effects due to chemical fractionation. In the following these two processes are described.

\vspace{-0.3cm}
\subsection{Nucleosynthesis processes}

\label{sec-isotope-galactic}

Isotopic ratios can be used as indicators of stellar nucleosynthesis. For example, $^{12}$C is purely a primary element, i.e., it forms from a mixture of H and He in the He-burning zones of stars of all masses. The main isotope of nitrogen, $^{14}$N, has a much more complex origin, being partly of primary and partly of secondary origin. In the latter case, its production needs the presence of $^{12}$C and $^{16}$O seeds already present at the star’s birth (\citealt{romano2017,romano2019}). 

The synergy between molecular observations and Galactic chemical evolution (GCE) models has imposed new important constraints onto the chemical evolution of the Milky Way. 
\citet{colzi18b} performed the first study of the $^{14}$N/$^{15}$N ratio of HCN and HNC as a function of the galactocentric distance, R$_{\rm GC}$, towards a large sample of $\sim$100 star-forming clumps observed with the IRAM 30m radiotelescope.
They found that the N-isotopic ratio increases from 2 up to 11 kpc, confirming what suggested by previous works (e.g. \citealt{adande2012}), and it decreases in the outer Galaxy (\citealt{colzi18a,colzi18b,colzi2022b}), with a new local ISM $^{14}$N/$^{15}$N value of 375$\pm$75. This observational trend has been compared with GCE models (\citealt{romano2017,romano2019}), and the production rate of elements and isotopes due to different kinds of stars across the Galaxy has been constrained.
In the outer Galaxy, accurate $^{14}$N stellar production rates in low-metallicity massive stars, which were previously unknown, have been determined. Moreover, the authors discovered that $^{15}$N has an important secondary production from nova outbursts produced in stellar binary systems in the inner Galaxy (R$_{\rm GC}$ $<$12 kpc).

Regarding carbon, several observations towards star-forming regions in the inner Galaxy have shown that the $^{12}$C/$^{13}$C ratio increases with R$_{\rm GC}$ (\citealt{milam2005,yan2019}, and references within). 
As for nitrogen, a Galactic gradient in the carbon isotopic ratio may be due to stellar nucleosynthesis processes (\citealt{romano2019,romano2021}). However, single-dish studies have limitations, including the large beam sizes, which sample less well-defined areas. Furthermore, only a handful of studies have extensive source number statistics ($N\geq50$, e.g. \citealt{colzi18b}), and, in general, with relatively large error bars. Recent work by Law et al. (2024, under consortium review) attempted to determine the carbon isotopic ratio as a function of R$_{\rm GC}$ using molecular line observations of high-mass star-forming regions from the ALMA Evolutionary study of High Mass Protocluster Formation in the Galaxy (ALMAGAL) survey (ID: 2019.1.00195.L; PIs: Molinari, Schilke, Battersby, Ho). In contrast to many previous single-dish studies, the study reveals a lack of a radial carbon isotopic ratio gradient, which can be interpreted as evidence for local fractionation processes (e.g. \citealt{colzi2020}).
However, the study also discusses the possibility that optical depth plays a crucial role in biasing systematically the carbon isotope ratio, making the interpretation difficult. The main results establish that the isotopic ratio measurement requires an accurate multi-line analysis to correct for optical depth and to properly determine the excitation temperature. In practice, joint single-dish and interferometric observations would provide a way out. 
Other possible ways out include studies of the isotopic ratio with heavier species, which are optically thinner, and doubly substituted isotopologues (e.g. \citealt{tercero2024}), which require longer integration times, making survey-like observations difficult.

\vspace{-0.3cm}
\subsection{Local nitrogen fractionation}
\label{sec-isotope-frac}

As mentioned before, molecular isotopic ratios are also governed by local fractionation processes. Low-temperature isotopic-exchange reactions are able to explain the D/H ratios (e.g. \citealt{caselliceccarelli2012}), but they do not reproduce the observed $^{14}$N/$^{15}$N ratios (e.g. \citealt{roueff2015,loison2019,sipila2023}). Recent works have shown that other chemical processes could play an important role: (i)  isotope selective photodissociation of N$_{2}$ (e.g. \citealt{furuya2018}), and (ii) different rates for the dissociative recombination of N$_{2}$H$^{+}$ (e.g. \citealt{loison2019}).

The effects of the first process have been investigated by \citet{colzi2019}, who  studied for the first time N-fractionation of N$_{2}$H$^{+}$ at high spatial resolution ($\sim$0.03 pc) towards the massive star-forming protocluster IRAS 05358+3543. They found a lower $^{14}$N/$^{15}$N ratio of N$_{2}$H$^{+}$ in the inner denser cores ($\sim$0.03 pc) of the cluster (100--200) with respect to the more diffuse gas ($\sim$0.5 pc) in which the cores are embedded ($>$250), in agreement with the prediction of the chemical model of \citet{furuya2018}. In the more diffuse part of a molecular clump, where the external UV radiation field is not fully shielded, the $^{14}$N$^{15}$N molecule could be photodissociated where the N$_{2}$ is not (thanks to $^{14}$N$_{2}$ being more abundant and thus easier to self-shield), causing an increase of the N$_{2}$/$^{14}$N$^{15}$N ratio. Since N$_{2}$H$^{+}$ is a daughter molecule of N$_{2}$, the $^{14}$N/$^{15}$N ratio of N$_{2}$H$^{+}$ follows the N$_{2}$ ratio and increases. 

In the low-mass regime, there is also growing evidence of the importance of selective photodissociation on the N isotopic ratio. Both \citet{spezzano2022} and \citet{Redaelli2023} reported a spatial gradient of the $^{14}$N/$^{15}$N ratio in HCN and NH$_3$, respectively, towards the prestellar core L1544 ($\sim$0.02 pc of resolution). In particular, the ratio in the two molecules presents similar values and decreases towards the outskirts of the core, which is more exposed to the interstellar UV field. At the same time, at the source's center the $^{14}$NH$_3$/$^{15}$NH$_3$ is a factor of two lower than that of N$_{2}$H$^{+}$, which reaches $^{14}$N/$^{15}$N$\sim$1000. \cite{redaelli2018, redaelli2020} showed that N$_{2}$H$^{+}$ presents anti-fractionated behaviour towards cold and dense pre-stellar sources, whilst its isotopic ratio is consistent with the elemental value of the local ISM towards protostellar cores, where the feedback heats the gas and causes the thermal desorption of CO. These pieces of evidence suggest that, beyond selective photodissociation, isotope-dependency of N$_{2}$H$^{+}$ dissociative recombination (DR) might also play a role. This would alter the N$_{2}$H$^{+}$/$^{15}$N$_{2}$H$^{+}$ ratio in the cold and dense environments where DR is the dominant pathway of destruction for this species. No dedicated laboratory study is available for the ISM physical conditions, but at high temperatures (300$\,$K) these reaction rates are known to be isotope-dependent \citep{lawson2011}.

\vspace{-0.3cm}
\section{The importance of cosmic rays}
\label{sec-cr}

The interpretation of observed molecular abundances necessarily involves the use of astrochemical models. The latter are based on massive chemical networks (e.g. KIDA, UMIST) that consider the different reaction mechanisms between neutrals, charged particles, dust grains as well as photo- and cosmic-ray processes.
As soon as the H$_{2}$ column density is larger than about $10^{21}~{\rm cm}^{-2}$, the UV photons of the interstellar radiation field are completely absorbed and cosmic rays drive the processes of ionisation, excitation, dissociation and heating \citep[see e.g.][]{Padovani2020,Padovani2022,Padovani2024}.
Cosmic rays with energy below 1~GeV can ionise both atomic and molecular hydrogen, to finally form the trihydrogen cation H$_{3}^{+}$. This initiates a series of reactions that lead to the formation of more and more complex species up to prebiotic molecules.

Several observational techniques at radio and infrared frequencies \citep[e.g.][]{Indriolo2012,Bialy2020,Bovino2020} provide an estimate of the spectrum of low-energy cosmic rays in interstellar clouds by determining the cosmic-ray ionisation rate, $\zeta$, which is the main parameter of astrochemical codes. Observations indicate that the assumption of a constant $\zeta$, derived from the foundational research of L. Spitzer in the 1950s, is not accurate. Indeed, $\zeta$ exhibits a significant variation, spanning at least 4$-$5 orders of magnitude, ranging from $10^{-15}\ {\rm s}^{-1}$ in diffuse clouds \citep{Luo2023} to $10^{-19}\ {\rm s}^{-1}$ in circumstellar-disc midplanes \citep{Cleeves2018}. Resolved maps of $\zeta$ are now also available \citep{Sabatini2023,Pineda2024}, showing potential gradients of the ionisation fraction through star-forming regions. In addition, higher rates have been estimated in protostellar sources (e.g. \citealt{ceccarelli2014b,Fontani2017}) and at the surface of protostellar shocks \citep[e.g.][]{Podio2014,Lattanzi2023}, up to a few $10^{-14}$~s$^{-1}$, which cannot be attributed to the Galactic cosmic ray flux.
All this poses a crucial warning for modellers of astrochemical codes and non-ideal magnetohydrodynamical simulations.

\begin{acknowledgements}
 L.C. and V.M.R. acknowledge funding from grants No. PID2019-105552RB-C41 and PID2022-136814NB-I00 by the Spanish Ministry of Science, Innovation and Universities/State Agency of Research MICIU/AEI/10.13039/501100011033 and by ERDF, UE. V.M.R. also acknowledges support from the grant number RYC2020-029387-I funded by MICIU/AEI/10.13039/501100011033 and by "ESF, Investing in your future", and from the Consejo Superior de Investigaciones Cient{\'i}ficas (CSIC) and the Centro de Astrobiolog{\'i}a (CAB) through the project 20225AT015 (Proyectos intramurales especiales del CSIC).
\end{acknowledgements}

%



\bibliographystyle{aa}
\bibliography{bibliography}

\end{document}